\documentclass[letterpaper]{article} 
\usepackage{aaai2026}  
\usepackage{times}  
\usepackage{helvet}  
\usepackage{courier}  
\usepackage[hyphens]{url}  
\usepackage{graphicx} 
\usepackage[numbers]{natbib}  
\usepackage{caption} 
\usepackage{algorithm}
\usepackage{algorithmic}
\usepackage{enumitem}
\usepackage{amssymb}
\usepackage{amsmath}
\usepackage{multirow,multicol}
\usepackage{subfigure}
\usepackage{booktabs}
\usepackage{xcolor}
\usepackage{makecell}
\newcommand{\name}{\textit{RouteMark}} 

\usepackage{newfloat}
\usepackage{listings}
\DeclareCaptionStyle{ruled}{labelfont=normalfont,labelsep=colon,strut=off} 
\floatstyle{ruled}
\newfloat{listing}{tb}{lst}{}
\floatname{listing}{Listing}
\title{\name: A Fingerprint for Intellectual Property Attribution in Routing-based Model Merging}
\author{
    Xin He\textsuperscript{\rm 1}\thanks{Corresponding author: hex@cfar.a-star.edu.sg},
    Junxi Shen\textsuperscript{\rm 1},
    Zhenheng Tang\textsuperscript{\rm 2},
    Xiaowen Chu \textsuperscript{\rm 3},
    Bo Li\textsuperscript{\rm 2},
    Ivor W. Tsang\textsuperscript{\rm 1},
    Yew-Soon Ong\textsuperscript{\rm 1}\textsuperscript{\rm 4}
}
\affiliations{
    \textsuperscript{\rm 1} CFAR and IHPC, Agency for Science, Technology and Research (A*STAR), \\
    \textsuperscript{\rm 2} Hong Kong Univerisity of Science and Technology \\
    \textsuperscript{\rm 3} The Hong Kong University of Science and Technology (Guangzhou) \\
    \textsuperscript{\rm 4} Nanyang Technological University (NTU), Singapore 
}

\usepackage{bibentry}

\begin{document}

\maketitle

\begin{abstract}

Model merging via Mixture-of-Experts (MoE) has emerged as a scalable solution for consolidating multiple task-specific models into a unified sparse architecture, 
where each expert is derived from a model fine-tuned on a distinct task. 
While effective for multi-task integration, this paradigm introduces a critical yet underexplored challenge: 
how to attribute and protect the intellectual property (IP) of individual experts after merging.  
We propose \textbf{\name}, a framework for IP protection in merged MoE models through the design of \emph{expert routing fingerprints}.  
Our key insight is that task-specific experts exhibit stable and distinctive routing behaviors under probing inputs.  
To capture these patterns, we construct expert-level fingerprints using two complementary statistics: 
the \textit{Routing Score Fingerprint (RSF)}, quantifying the intensity of expert activation, 
and the \textit{Routing Preference Fingerprint (RPF)}, characterizing the input distribution that preferentially activates each expert.  
These fingerprints are reproducible, task-discriminative, and lightweight to construct.  
For attribution and tampering detection, we introduce a similarity-based matching algorithm that compares expert fingerprints between a suspect and a reference (victim) model.  
Extensive experiments across diverse tasks and CLIP-based MoE architectures show that \textbf{\name} consistently yields high similarity for reused experts and clear separation from unrelated ones.  
Moreover, it remains robust against both \emph{structural tampering} (expert replacement, addition, deletion) and \emph{parametric tampering} (fine-tuning, pruning, permutation), 
outperforming weight- and activation-based baseliness. 
Our work lays the foundation for \textbf{\name} as a practical and broadly applicable framework for IP verification in MoE-based model merging.

\end{abstract}

\section{Introduction}

Large pretrained models such as GPT~\cite{gpt1, gpt2, gpt3}, T5~\cite{t5}, and CLIP~\cite{radford2021CLIP} have become foundation models across a wide range of AI applications. While these models provide a strong backbone, enabling them to perform well across diverse tasks still typically requires task-specific fine-tuning, which can be computationally expensive when scaled to many tasks. To address this challenge, model merging~\cite{yang2024modelmergingSurvey} has emerged as a cost-effective alternative: instead of retraining, one can directly integrate multiple fine-tuned models into a unified system, allowing shared deployment and multi-task capability with minimal additional overhead.

\begin{figure}[!htb]
    \centering
    \includegraphics[width=\linewidth]{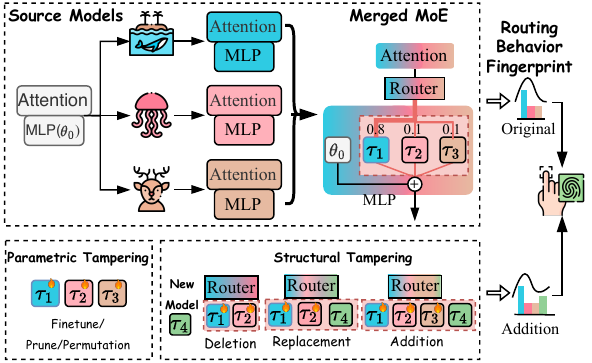}
    
    \caption{Overview of our routing-based fingerprinting framework \textit{\name}.
We merge task-specific experts into an MoE via a learned router. Instead of relying on model weights or activations, our \textit{\name} constructs expert fingerprints based on routing logits. It detects unauthorized parametric and structural tampering operations by capturing consistent expert-specific routing behaviors.}
    \label{fig:overview}
\end{figure}

However, the growing adoption of model merging raises serious concerns about intellectual property (IP) attribution and protection. As multiple fine-tuned models are blended into a single model, it becomes increasingly difficult to verify whether a component model has been reused. Recent studies have begun to explore this issue under static merging strategies such as parameter averaging~\cite{wortsman2022modelsoup}, task arithmetic~\cite{ilharco2023editingmodels}, and Ties-Merging~\cite{yadav2023tiesMerging}. These works show that watermarking methods~\cite{li2023quantwatermarking} often fail to survive the merging process, and fingerprinting techniques~\cite{xu2024instructionalfingerprint} exhibit high sensitivity to merge configurations, particularly the merge ratio. Even small changes in blending weights can significantly weaken the fingerprint signal. This limitation prompts a broader question: \textit{how can IP attribution be performed reliably in more dynamic and input-dependent merging scenarios, where model behavior varies across inputs and attribution becomes harder to disentangle?}

Motivated by the scalability and conditional computation of Mixture-of-Experts (MoE) architectures~\cite{shazeer2017outrageouslyMoE, fedus2022switchtransformer,he2024expertflow,rajbhandari2022deepspeedmoe}, 
recent works have explored adopting MoE as a basis for model merging~\cite{tang2024mergingMoE, shen2024MoEMergeMulTask, kang2024selfMoE, yadav2024surveyMoEMerge}.  
In this setting, each expert module is typically derived from a fine-tuned model trained on a specific task, and a learned router assigns routing scores to select a small subset of experts for each input.  
This dynamic activation mechanism enables the merged model to scale efficiently across heterogeneous tasks while maintaining modular specialization.  
However, it also makes attribution significantly more challenging, since expert contributions vary across inputs and layers, 
obscuring the influence of individual experts.

Moreover, the modular structure of MoE allows for fine-grained tampering. An adversary may fine-tune experts to obfuscate their identity, inject redundant experts to confuse attribution, remove or replace original experts, or permute expert parameters to disrupt alignment. These operations pose major challenges for existing fingerprinting methods. For instance, weight-based approaches such as \textit{PCS} and \textit{ICS}~\cite{zeng2024huref} rely on parameter similarity but are easily defeated by simple permutations. Activation-based approaches like \textit{REEF}~\cite{zhang2024reefCKA} compare internal layer representations but fail to pinpoint which expert has been reused or tampered with. Their coarse granularity and sensitivity to task shifts further limit practical utility under MoE-based merging.

To this end, we propose \textit{\name}, a fingerprinting framework that captures each expert’s behavioral signature through routing logits observed during inference.  
As illustrated in Fig.~\ref{fig:overview}, \textit{\name} constructs expert-specific fingerprints using routing information collected on a fixed set of probing inputs, 
without relying on model weights or internal activations.  
It integrates two complementary components: (1) the \textit{Routing Score Fingerprint (RSF)}, quantifying each expert’s activation intensity across tasks and layers, 
and (2) the \textit{Routing Preference Fingerprint (RPF)}, characterizing its relative selection tendencies and task-level specialization.  
Extensive experiments show that \textit{\name} enables reliable expert attribution under diverse tampering scenarios.  
Our main contributions are as follows:

\begin{enumerate}
    \item \textbf{Problem Formulation}. We provide the first systematic study of IP attribution under MoE-based model merging, clarifying its unique challenges and threat model.
    
    \item \textbf{Routing-based Fingerprinting}. We introduce \textit{\name}, a novel method that leverages routing behavior via two complementary descriptors, \textit{RSF} and \textit{RPF}, jointly capturing activation intensity and selection preference.
    
    \item \textbf{Robustness to Tampering}. We investigate a diverse set of structural and parametric tampering operations, including fine-tuning, replacement, addition, deletion, pruning, and permutation, and demonstrate that \textit{\name} consistently outperforms weight- and activation-based baselines.
\end{enumerate}

\section{Related Work}

\subsection{MoE Architectures}

MoE architectures~\cite{shazeer2017outrageouslyMoE, fedus2022switchtransformer, rajbhandari2022deepspeedmoe} effectively increase model capacity without proportionally raising computational demands. By employing a learned routing function to dynamically select a subset of experts for each input, MoEs achieve conditional computation by activating only relevant model components. This input-adaptive selection promotes expert specialization and modular knowledge reuse, particularly beneficial for integrating heterogeneous sources~\cite{he2024expertflow,xue2024openmoe}. Consequently, MoEs enhance computational efficiency and task-specific performance in multi-task scenarios.

\subsection{Model Merging}

Model merging~\cite{yang2024modelmergingSurvey} consolidates multiple pretrained or fine-tuned models into a unified system, facilitating efficient deployment and knowledge reuse without extensive retraining. Early approaches mainly merged dense models through weight averaging~\cite{wortsman2022modelsoup, yadav2023tiesMerging}, task arithmetic~\cite{ilharco2023editingmodels}, or representation alignment~\cite{stoica2023zipit}. However, these static methods, which rely on fixed merging ratios and shared architectures, lack adaptability to varying inputs and frequently experience performance degradation from conflicting objectives.

Recent studies propose modular merging frameworks, particularly MoE-based approaches~\cite{tang2024mergingMoE, shen2024MoEMergeMulTask, yadav2024surveyMoEMerge, kang2024selfMoE, li2023mergeCompress, lu2024twinMerge, muqeeth2023softMerge}. In MoE merging, each fine-tuned model serves as an independent expert, and a learned router dynamically selects relevant experts based on the input. This approach offers adaptive expert composition, eliminates manual hyperparameter tuning, preserves expert specialization, and enhances scalability and generalization in multi-task scenarios.

\begin{table}[t]
\centering
\caption{Comparison of fingerprinting methods. ``Finetune‑Free'' indicates no extra training needed for fingerprint generation;  
``Probe‑Efficient'' means requiring no or only a small fixed probe set;  
``Granularity'' specifies whether attribution is at model or expert level.}
\label{tab:fingerprint_comparison}
\scalebox{0.9}{
\begin{tabular}{l|c|c|c|c}
\hline
\textbf{Method} &\makecell{\textbf{Signal}\\\textbf{Source}} & \makecell{\textbf{Finetune-}\\\textbf{Free}} & \makecell{\textbf{Probe-}\\ \textbf{Efficient}} & \makecell{\textbf{Granul-}\\\textbf{arity}} \\
\hline
KGW~\cite{redGreenWatermark} & Logits & \checkmark & \checkmark & Model \\
QW~\cite{li2023quantwatermarking} & Weights & $\times$ & $\times$ & Model \\
IF~\cite{xu2024instructionalfingerprint} & Outputs & $\times$ & $\times$ & Model \\
REEF~\cite{zhang2024reefCKA} & Activations & \checkmark & \checkmark & Model \\
PCS~\cite{zeng2024huref} & Weights & \checkmark & \checkmark & Model \\
ICS~\cite{zeng2024huref} & Weights & \checkmark & \checkmark & Model \\
\hline
Ours & Routing & \checkmark & \checkmark & Expert \\
\hline
\end{tabular}
}
\end{table}


\subsection{Model Fingerprinting}

Model fingerprinting embeds a unique identity into a trained model for ownership verification, provenance tracking, and unauthorized usage detection.  
As summarized in Table~\ref{tab:fingerprint_comparison}, prior approaches exploit diverse signals.  
Logit-based watermarks such as KGW~\cite{redGreenWatermark} inject special patterns into predictions but are fragile to post-hoc modifications.  
Weight-based schemes including QW~\cite{li2023quantwatermarking}, PCS, and ICS~\cite{zeng2024huref} compare parameter statistics, 
achieving discriminability yet remaining highly sensitive to reordering or tampering.  
Instruction-conditioned methods like IF~\cite{xu2024instructionalfingerprint} embed behavioral triggers but require carefully designed prompts and substantial data.  
Activation-based techniques such as REEF~\cite{zhang2024reefCKA} use representation similarity via the Centered Kernel Alignment (CKA) metric~\cite{CKA}, 
offering partial robustness to reparameterization.  
While some are finetune-free or data-efficient, none provide reliable expert-level attribution.  

Fingerprint robustness under model merging has also been studied.  
MergePrint~\cite{Yamabe2024MergePrintRF} shows that conventional fingerprints often lose detectability once models are blended.  
Cong et al.~\cite{cong2024HaveyouMerged} benchmark QW and IF under dense static merging methods, 
demonstrating that even minor variations in merge ratios can substantially degrade fingerprint signals.  
However, these works remain limited to static merging, leaving the challenge of expert attribution in modular, dynamic MoE-based merging unexplored.  
To fill this gap, our \textit{\name} leverages expert routing behaviors to generate finetune-free, data-efficient fingerprints, 
and uniquely enables robust expert-level attribution in MoE architectures.

\section{Preliminaries}

\subsection{MoE-Based Model Merging}\label{sec:moe_merging}

Given a pretrained foundation model $\mathcal{M}_0$ with parameters $\theta_0$, we consider $N$ task-specific fine-tuned models $\{\mathcal{M}_1, \dots, \mathcal{M}_N\}$, each trained on a dataset $\mathcal{D}_i$ to obtain parameters $\theta_i$. For each task, we compute its task vector (or delta) as:
\begin{equation}
    \Delta_i = \theta_i - \theta_0.
\end{equation}

These task vectors are used as experts and then merged into a unified model $\mathcal{V}$, which is referred to as the \textbf{victim model} and follows an MoE-style architecture~\cite{tang2024mergingMoE}. We adopt distinct strategies for attention and MLP modules.

\paragraph{Attention}
For attention layers, we use uniform averaging:
\begin{equation}
    \theta^\text{attn}_\text{merge} = \theta_0^\text{attn} + \lambda \cdot \sum_{i=1}^{N} \Delta_i^\text{attn},
\end{equation}
where $\lambda$ is a fixed scalar controlling the merge strength.

\paragraph{MLP}
For MLP blocks, we apply a task-weighted  fusion:
\begin{equation}
    \theta^\text{mlp}_\text{merge} = \theta_0^\text{mlp} + \sum_{i=1}^{N} \alpha_i \cdot \Delta_i^\text{mlp},
\end{equation}
where $\alpha_i \in [0, 1]$ are task-specific fusion weights satisfying $\sum_{i=1}^N \alpha_i = 1$. These weights are dynamically generated by a learned router $\mathcal{R}_\text{mlp}$ from a task vector $z$:
\begin{equation}
    \alpha = \mathcal{R}_\text{mlp}(z), \quad \text{with } \alpha_i = \frac{\exp(g_i(z))}{\sum_{j=1}^N \exp(g_j(z))},
\end{equation}
where $g_i(z)$ is the relevance score of task $i$.

Once trained, $\mathcal{R}_\text{mlp}$ outputs a global fusion vector $\alpha$, resulting in a shared $\theta^\text{mlp}_\text{merge}$ across all tokens, contrasting with traditional token-level MoE activation. This design yields a compositional and parameter-efficient architecture for multi-task deployment.

\subsection{Tampering Behaviors in Merged MoE}
\label{sec:tampering}

The increasing openness of foundation models has made it common to release not only the pretrained backbone $\mathcal{M}_0$, but also fine-tuned variants $\{\mathcal{M}_i\}$ and their merged MoE model $\mathcal{V}$. While this modularity encourages reusability, it also creates new risks of unauthorized reuse. In particular, the expert deltas $\{\Delta_i\}$ encapsulate task-specific knowledge in a plug-and-play manner, enabling fine-grained tampering without changing the overall architecture. 

A suspect model $\mathcal{S}$ can thus preserve the high-level performance of $\mathcal{V}$ while covertly reusing or manipulating its experts. We categorize common tampering strategies into two types: \textit{Structural Tampering}, which changes the composition of the expert pool, and \textit{Parametric Tampering}, which alters the internal parameters of existing experts.

\paragraph{Structural Tampering.}
These operations directly modify the expert pool by adding, removing, or replacing experts:
\begin{itemize}
    \item \textbf{Expert Replacement.} One or more original experts $\Delta_j$ are substituted with tampered versions $\Delta'_j$:
    \begin{align}
    \theta^\text{mlp}_\mathcal{S} = \theta^\text{mlp}_0 + \sum_{i \ne j} \alpha_i \cdot \Delta_i + \alpha_j \cdot \Delta'_j.
    \end{align}
    \item \textbf{Expert Addition.} New deltas $\{\Delta_{N+1}, \dots, \Delta_{N'}\}$ are added to the pool:
    \begin{align}
    \theta^\text{mlp}_\mathcal{S} = \theta^\text{mlp}_0 + \sum_{i=1}^{N'} \tilde{\alpha}_i \cdot \Delta_i, \quad N' > N.
    \end{align}
    \item \textbf{Expert Deletion.} Only a subset $\mathcal{I}$ of the original experts is reused, with routing weights retrained:
    \begin{align}
    \theta^\text{mlp}_\mathcal{S} = \theta^\text{mlp}_0 + \sum_{i \in \mathcal{I}} \bar{\alpha}_i \cdot \Delta_i,\quad \mathcal{I} \subsetneq \{1, \dots, N\}.
    \end{align}
\end{itemize}

\paragraph{Parametric Tampering.}
These operations manipulate the internal parameters of existing experts while keeping the expert pool intact:
\begin{itemize}
    \item \textbf{Expert Finetuning.} The original experts are further trained to adjust expert behavior, while maintaining the overall architecture and output performance.
    \item \textbf{Expert Pruning.} Low-importance weights are removed to introduce sparsity, a technique often applied for  compression, which may also affect attribution.
    \item \textbf{Expert Permutation.} The hidden dimensions within an expert are permuted, disrupting parameter alignment without altering functional behavior~\cite{fernandez2024functional}.  
\end{itemize}


\begin{figure*}[!htb]
    \centering
    \subfigure[t-SNE visualization]{\includegraphics[width=0.25\textwidth]{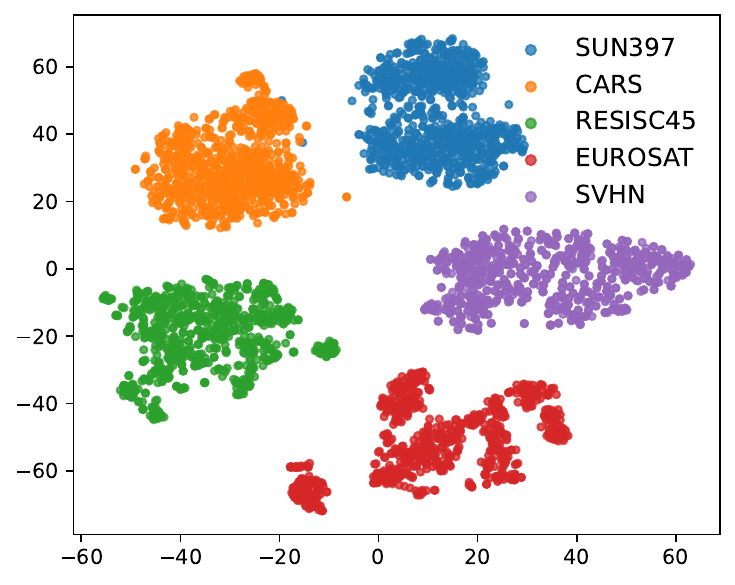}
    }
    \subfigure[Expert routing patterns ]{    \includegraphics[width=0.73\textwidth]{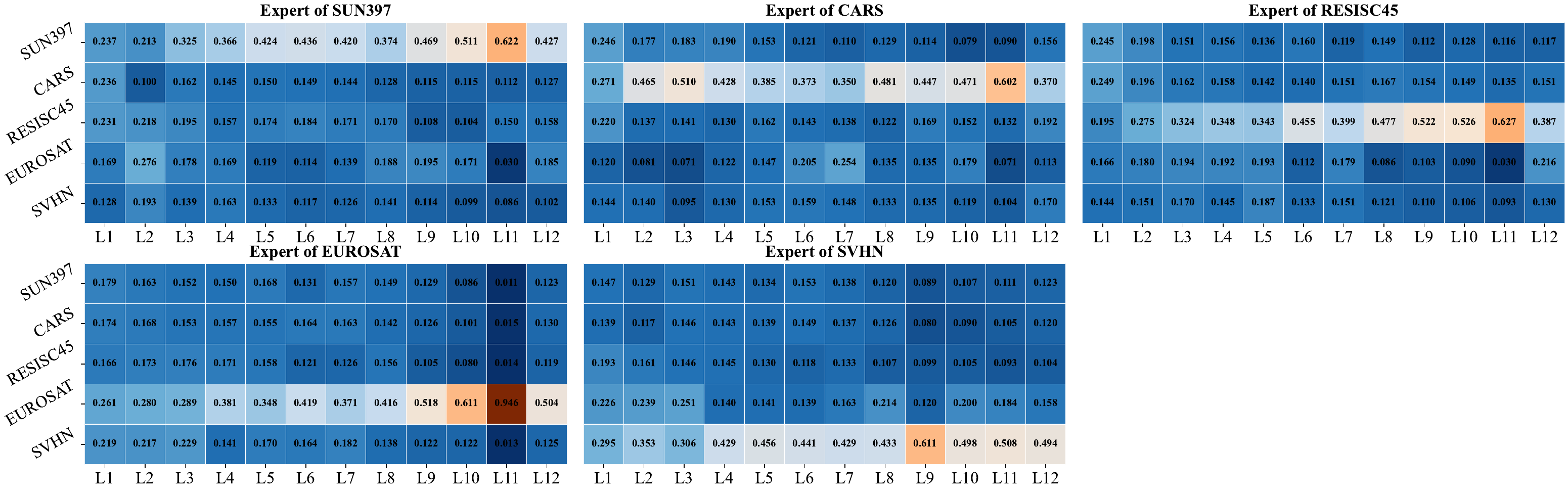}
    }
    \caption{Task-aware behavior in the MoE merged from five expert models. (a) Token embeddings form well-separated clusters by dataset, indicating preserved semantic distinctions. (b) Heatmaps show the routing weights for individual experts across MoE layers. Each expert is most active on its original training dataset, revealing task-aligned routing behavior.}
    
    \label{fig:pre-method}
\end{figure*}

\section{Method}
\label{sec:method}

This section presents our fingerprint framework \textit{\name} for detecting expert reuse in MoE-based model merging. We begin with empirical evidence showing that merged experts maintain task-consistent routing behaviors. Based on this, we formalize expert fingerprints derived from aggregated routing statistics. Finally, we detail our similarity-based matching algorithm, enabling expert-level attribution and remains robust under diverse tampering operations.

\subsection{Task-Aligned Routing as Fingerprint Signals}

The experts within the merged MoE often exhibit input-dependent routing patterns that reflect the original tasks of their constituent experts. To examine this, we analyze the token representations and routing behavior of a merged model $\mathcal{V}$ composed of five task-specific experts.

As shown in Fig.~\ref{fig:pre-method}(a), token embeddings from different tasks form well-separated clusters in the latent space, indicating preserved semantic distinctions. Building on this, Fig.~\ref{fig:pre-method}(b) presents expert-wise heatmaps of average routing weights across layers. Each expert shows peak activation on its original training dataset, demonstrating consistent task-aligned routing behavior.

These results suggest that each expert retains a distinctive activation profile across tasks. We leverage this property by constructing compact \emph{expert routing fingerprints} based on aggregated routing statistics, enabling expert-level attribution without relying on output or weight comparisons.

\subsection{Expert Routing Fingerprint Construction}
\label{sec:fingerprint}

We characterize each expert by its routing behavior across tasks and mixture layers.  
For expert $j$, its fingerprint is defined by two complementary components, both constructed from the internal routing logits generated by the MoE router when the merged model is evaluated on $N$ probe datasets $\{\mathcal{D}_1, \dots, \mathcal{D}_N\}$.  
These probe datasets correspond directly to the fine-tuning domains of the victim MoE’s experts and remain fixed throughout all experiments, 
ensuring consistent fingerprint construction for both victim and suspect models.

\begin{itemize}
  \item \textbf{Routing Score Fingerprint (RSF)}: a task-layer matrix capturing routing intensity, denoted as $F_j^{\mathrm{RSF}} \in \mathbb{R}^{N \times L}$;
  \item \textbf{Routing Preference Fingerprint (RPF)}: a task-level probability vector summarizing specialization, denoted as $F_j^{\mathrm{RPF}} \in \mathbb{R}^{N}$.
\end{itemize}

\subsubsection{Routing Score Fingerprint (RSF)}

Let the MoE model contain $E$ experts and $L$ mixture layers. For each input $x \in \mathcal{D}_i$, let $\alpha_j^{(l)}(x)$ denote the routing logit assigned to expert $j$ at layer $l$.

We compute the average routing logit across each dataset:
\begin{align}
F_{\mathrm{moe}}[i, j, l] = \mathbb{E}_{x \sim \mathcal{D}_i} \left[ \alpha_j^{(l)}(x) \right].
\end{align}

To remove layer-wise biases, we apply expert-wise mean-centering:
\begin{align}
\tilde{F}_{\mathrm{moe}}[i, j, l] = F_{\mathrm{moe}}[i, j, l] - \frac{1}{E} \sum_{j'=1}^{E} F_{\mathrm{moe}}[i, j', l].
\end{align}

The Routing Score Fingerprint for expert $j$ is extracted as the task-layer slice:
\begin{align}
F_j^{\mathrm{RSF}} = \tilde{F}_{\mathrm{moe}}[:, j, :] \in \mathbb{R}^{N \times L}.
\end{align}

\textit{RSF} offers a fine-grained view of the expert’s routing behavior across tasks and  layers, where each entry $F_j^{\mathrm{RSF}}[i, l]$ captures the activation intensity of expert $j$ on task $\mathcal{D}_i$ at layer $l$. It captures how intensely expert $j$ is activated across different tasks and layers, reflecting its input-dependent behavior under varying contexts. This structured matrix preserves layer-wise variation and serves as a fine-grained behavioral signature for expert alignment.

\subsubsection{Routing Preference Fingerprint (RPF)}

To summarize task preference, we aggregate $F_j^{\mathrm{RSF}}$ over mixture layers:
\begin{align}
s_j[i] = \sum_{l=1}^{L} F_j^{\mathrm{RSF}}[i, l], \quad s_j \in \mathbb{R}^{N}.
\end{align}

We then apply a softmax to obtain a task-level probability distribution:
\begin{align}
F_j^{\mathrm{RPF}}[i] = \frac{\exp(s_j[i])}{\sum_{i'=1}^{N} \exp(s_j[i'])}, \quad F_j^{\mathrm{RPF}} \in \mathbb{R}^{N}.
\end{align}

The resulting \textit{RPF} provides a task-level profile of each expert by forming a probability distribution over probe tasks.  
Because fingerprints are evaluated on fixed task-specific probes, victim experts show sharply peaked distributions on their respective target tasks, even after being subjected to diverse tampering operations.  
In contrast, new experts introduced through addition or replacement lack such alignment and therefore produce relatively uniform profiles across tasks.  
As confirmed in our experiments (\S\ref{sec:effectiveness}), this distinction enables \textit{RPF} to capture coarse-grained task affinity while naturally complementing the fine-grained activation selectivity encoded by \textit{RSF}.


\subsection{Fingerprint Similarity}
\label{sec:fingerprint-similarity}

Given two experts $j$ and $k$ from victim and suspect models, respectively, we compare their routing fingerprints based on both absolute activation structure and task preference behavior. Let $(F_j^{\mathrm{RSF}}, F_j^{\mathrm{RPF}})$ and $(F_k^{\mathrm{RSF}}, F_k^{\mathrm{RPF}})$ denote the fingerprint components of the two experts.

\paragraph{Score-based similarity.}
We compute the cosine similarity between their RSF, vectorized into row-major order:
\begin{equation}
\text{Sim}^{\mathrm{RSF}}(j, k) = \frac{1}{2}(1-\cos\left( F_j^{\mathrm{RSF}}, F_k^{\mathrm{RSF}}) \right).
\end{equation}

\paragraph{Preference-based similarity.}
We compute the Jensen-Shannon divergence between their RPF distributions across layers and convert it into a similarity score:
\begin{equation}
\text{Sim}^{\mathrm{RPF}}(j, k) = 1 - \frac{1}{L} \sum_{l=1}^{L} \text{JSD} \left( F_j^{\mathrm{RPF}}[:, l] \,\|\, F_k^{\mathrm{RPF}}[:, l] \right),
\end{equation}
where $\text{JSD}(p \,\|\, q)$ denotes the Jensen-Shannon divergence between two discrete distributions $p$ and $q$.



\paragraph{Final similarity score.}
The overall fingerprint similarity between a victim expert $j$ and a suspect expert $k$ is defined as the average of their two component similarities:
\begin{equation}
\text{Sim}(j, k) = \frac{1}{2} \left( \text{Sim}^{\mathrm{RSF}}(j, k) + \text{Sim}^{\mathrm{RPF}}(j, k) \right).
\end{equation}

To determine whether a suspect expert is reused from the victim model, 
we examine its similarity profile, defined as its similarities with all victim experts.  
If the profile shows a single dominant match, the expert is considered reused and attributed to the corresponding victim expert.  
If the profile is relatively uniform without a clear winner, the expert is treated as newly introduced.  
This criterion underlies our experimental analysis in \S\ref{sec:effectiveness}. 

\section{Experiment}

We conduct extensive experiments to evaluate the proposed routing-based fingerprinting framework, \textit{\name},  focusing on expert-level attribution, robustness under diverse tampering operations and  varying sample sizes.


\subsection{Experimental Settings}


\subsubsection{CLIP-based MoE Merging}

We build merged MoE models from task-specific experts, each obtained by fine-tuning CLIP~\cite{radford2021CLIP} on one of the following datasets: 
\texttt{SUN397}~\cite{SUN397}, 
\texttt{Cars}~\cite{Stanfordcars}, 
\texttt{RESISC45}~\cite{RESISC45}, 
\texttt{EuroSAT}~\cite{EuroSAT}, 
\texttt{SVHN}~\cite{SVHN}, 
\texttt{GTSRB}~\cite{GTSRB}, 
\texttt{MNIST}~\cite{MNIST}, and 
\texttt{DTD}~\cite{DTD}.  
These datasets span a broad spectrum of domains including natural scenes, aerial imagery, digits, traffic signs, and textures, providing a rigorous testbed for evaluating expert specialization under heterogeneous visual conditions.

Beyond constructing a standard merged MoE, we systematically vary the set of included experts to emulate potential tampering behaviors such as expert addition, replacement, or removal. This design reflects realistic attack scenarios in MoE-based merging, where adversaries may manipulate the expert set to obscure attribution or alter model functionality.

\subsubsection{Baseline Methods}

We compare against representative baselines at two levels of granularity, 
depending on whether they can identify  the specific experts that have been reused.  
\textbf{Model-level} methods operate on the merged model as a whole and detect global reuse, 
whereas \textbf{expert-level} methods attempt to attribute individual experts.

At the model level, we include three baselines.  
\textit{PCS}~\cite{zeng2024huref} flattens all weights into a single vector and computes cosine similarity. 
\textit{ICS}~\cite{zeng2024huref} builds invariant representations from the last layers for cosine similarity. 
\textit{REEF}~\cite{zhang2024reefCKA} compares activations using Centered Kernel Alignment (CKA)~\cite{CKA}.  
These methods effectively detect overall model reuse but cannot pinpoint specific experts.

At the expert level, \textit{PCS} and \textit{ICS} are extended to compare expert-specific delta parameters $\{\Delta_i\}$.  
In contrast, \textit{REEF} is infeasible here: after merging attention and MLP layers, 
the model no longer preserves distinct activation pathways of individual experts.  Our framework, \textit{\name}, also operates at the expert level but uses routing behavior instead of weights or activations, 
providing a more robust and interpretable basis for attribution.  
For clarity, we denote model-level \textit{PCS} and \textit{ICS} as \textit{PCS-M} and \textit{ICS-M}, 
and their expert-level variants as \textit{PCS-E} and \textit{ICS-E}.

\subsection{Effectiveness of Expert Attribution}\label{sec:effectiveness}

A reliable attribution method should satisfy that experts from same sources should exhibit high similarity, 
and those from different sources should remain clearly distinguishable with low cross-similarity. 
We evaluate whether \textit{\name} meets these conditions under diverse tampering scenarios, including expert deletion, addition, replacement, fine-tuning, and pruning.

\subsubsection{Settings}

The victim model $\mathcal{V}$ is constructed as a CLIP-based MoE composed of five task-specific experts, 
each fine-tuned independently on 
\texttt{SUN397}, 
\texttt{Cars}, 
\texttt{RESISC45}, 
\texttt{EuroSAT}, 
and \texttt{SVHN}.  
As shown in Fig.~\ref{fig:effectiveness-matrices}, we evaluate attribution across six settings:  
(a) the original victim-only model,  
(b) a suspect obtained by fine-tuning all victim experts,  
(c) pruning with WANDA~\cite{wanda} at 50\% sparsity,  
(d) deletion of the \texttt{SUN397} expert,  
(e) addition of two new experts trained on \texttt{GTSRB} and \texttt{MNIST},  
and (f) replacement of the \texttt{SUN397} expert with a \texttt{GTSRB} expert.

\begin{table*}[]
\centering
\caption{Similarity score comparison between model-level and expert-level baselines under parametric tampering.}
\begin{tabular}{l|rrr|rrr}
\toprule
\multirow{2}{*}{\makecell{\textbf{Parametric Tampering}}} & \multicolumn{3}{c|}{\textbf{Model-level}} & \multicolumn{3}{c}{\textbf{Expert-level}} \\ \cline{2-7} 
 & \textit{PCS-M} & \textit{ICS-M} & \textit{REEF} & \textit{PCS-E} & \textit{ICS-E} & Ours \\ 
\midrule
Prune-Magnitude (30\% sparsity) & 0.000 & 0.000 & 0.785 & 0.000 & 0.000 & \textbf{0.960} \\
Prune-WANDA (30\% sparsity) & 0.013 & 0.018 & 0.769 & 0.013 & 0.011 & \textbf{0.922} \\
Finetune (+4,000 steps) & 0.839 & 0.714 & 0.851 & 0.872 & 0.833 & \textbf{0.958} \\
Finetune (+8,000 steps) & 0.715 & 0.606 & 0.843 & 0.749 & 0.733 & \textbf{0.956} \\
Parameter Permutation & 0.014 & 0.574 & \textbf{1.000} & 0.001 & 0.000 & \textbf{1.000} \\ 
\bottomrule
\end{tabular}
\label{tab:parametric_tampering}
\end{table*}

\subsubsection{\textit{\name} attributes reused experts with high similarity and clear separation under diverse tampering}

\begin{figure}[!htb]
    \centering
    \subfigure[Original Victim]{
    \includegraphics[width=0.3\linewidth]{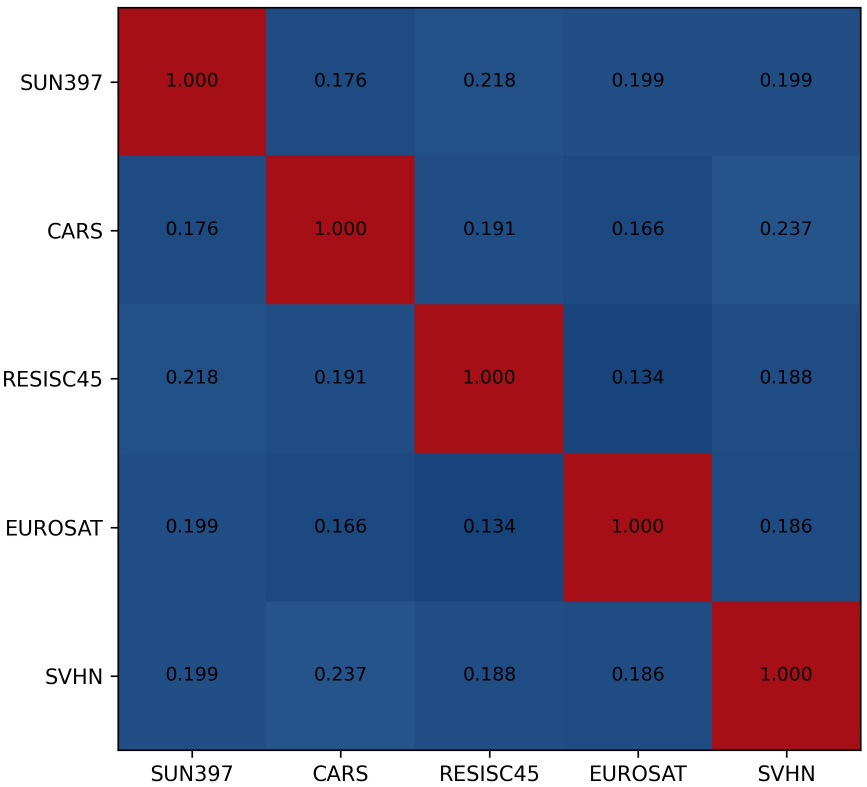}
    }
    \subfigure[Finetune Expert]{
    \includegraphics[width=0.3\linewidth]{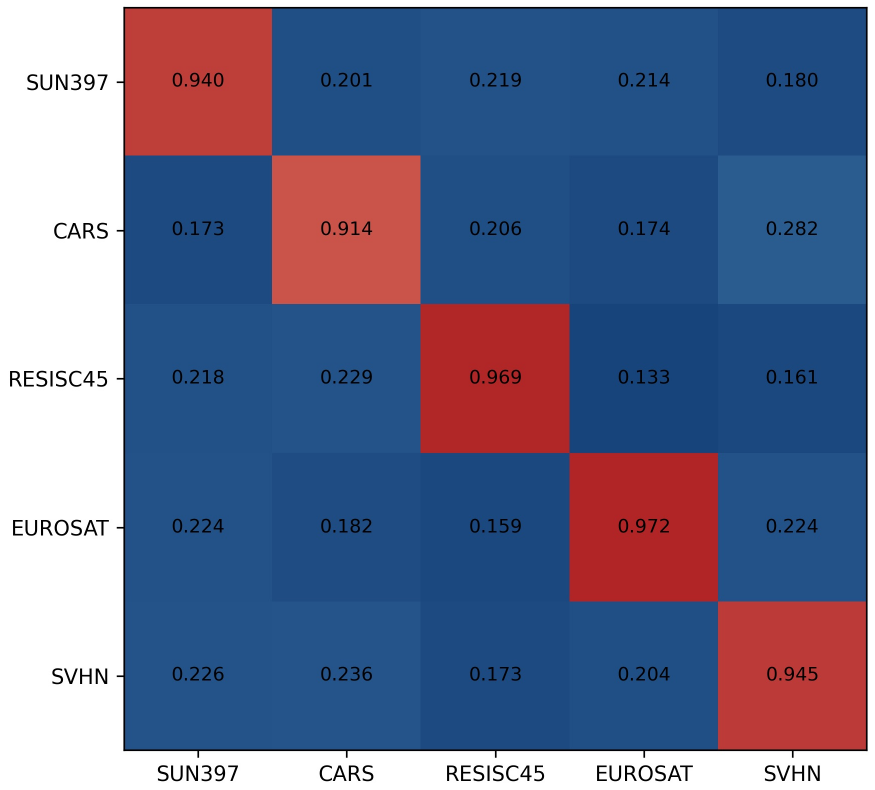}
    }
    \subfigure[Prune Expert]{
    \includegraphics[width=0.3\linewidth]{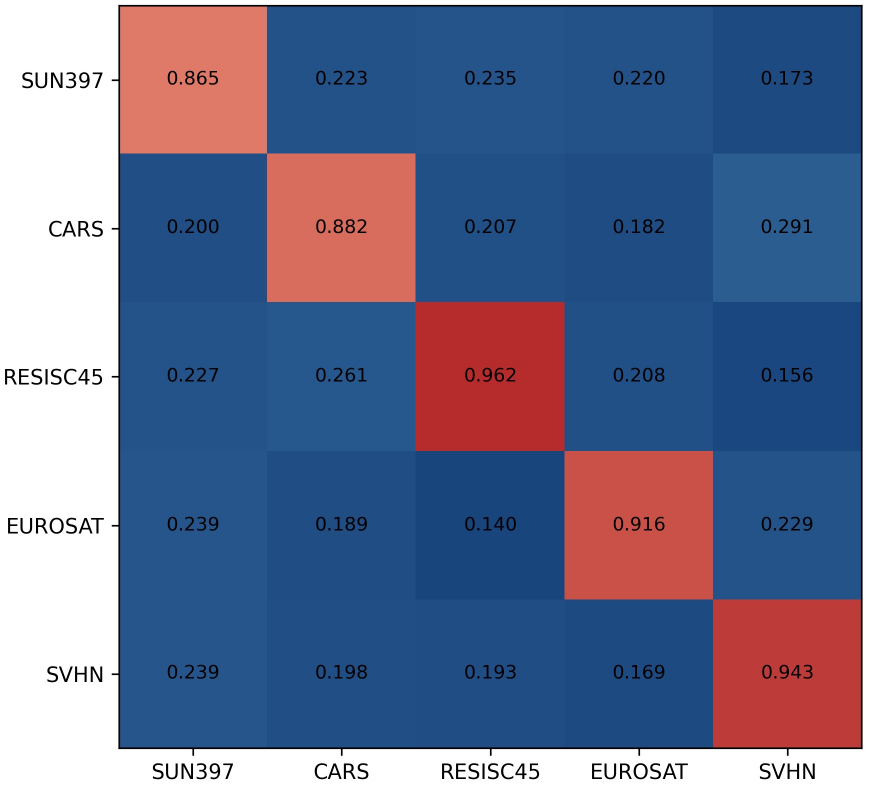}
    }
    
    \subfigure[Delete Expert]{
    \includegraphics[width=0.25\linewidth]{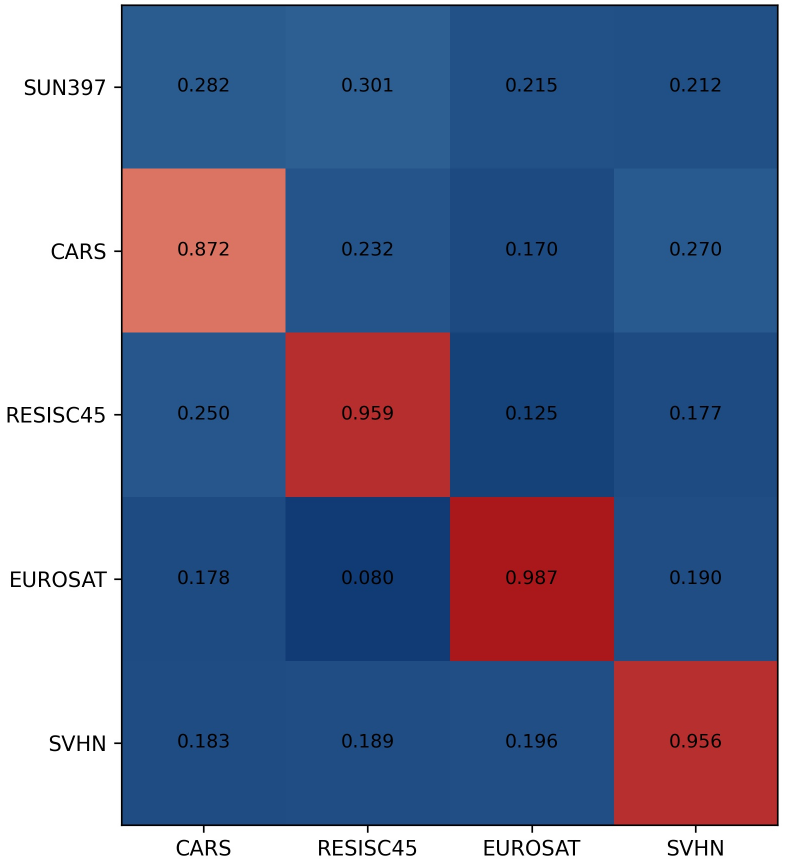}
    }
    \subfigure[Add Expert]{
    \includegraphics[width=0.35\linewidth]{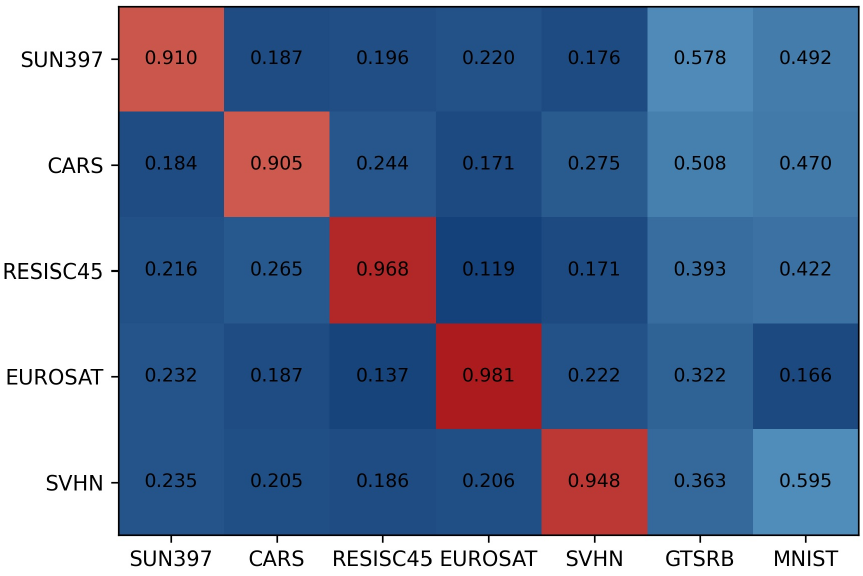}
    }
    \subfigure[Replace Expert]{
    \includegraphics[width=0.3\linewidth]{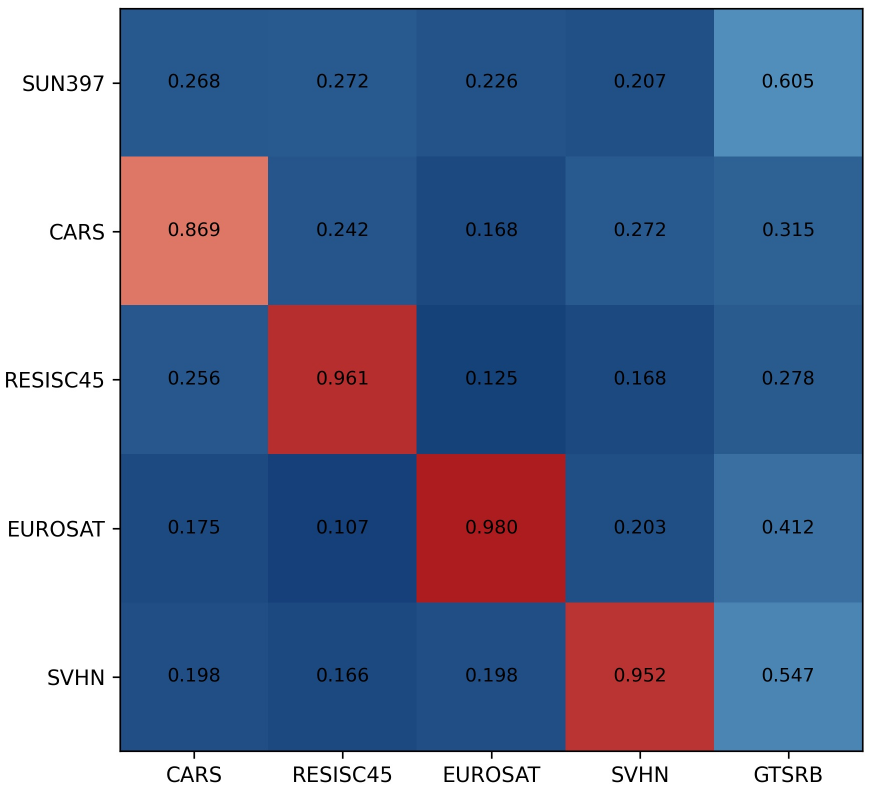}
    }
\caption{Fingerprint similarity matrices of (a) the original victim, (b–c) parametric tampering, and (d–f) structural tampering based on the fine-tuned suspect in (b). Since experts in (b–f) are fine-tuned, reused experts show diagonal similarities slightly below 1.0, yet their columns remain sharply peaked, ensuring clear identification of victim counterparts. In contrast, new experts introduced by addition and replacement yield more uniform distributions, indicating weak correspondence and avoiding misattribution.}
    \label{fig:effectiveness-matrices}
\end{figure}

Fig.~\ref{fig:effectiveness-matrices} presents fingerprint similarity matrices across six aforementioned cases. 
Each column corresponds to a suspect expert and its similarities to all victim experts.  
To assess attribution, we focus on whether columns exhibit a clear peak at the correct counterpart.  
We quantify this using two statistics: the \textit{Top-1 similarity}, i.e., the highest value within a column,  
and the \textit{similarity margin}, defined as the difference between the Top-1 and Top-2 values.  
A reliable method should produce both high Top-1 similarity for reused experts and large similarity margins to separate them from unrelated ones.

For reused experts, Fig.~\ref{fig:effectiveness-matrices}(a) first shows that experts within the victim model are inherently well separated,  
with most off-diagonal similarities below 0.20.  
Across tampering cases in Fig.~\ref{fig:effectiveness-matrices}(b–f), reused experts maintain strong alignment with their victim counterparts:  
their average similarities are 0.948, 0.914, 0.944, 0.942, and 0.941 under fine-tuning, pruning, deletion, addition, and replacement, respectively,  
with the maximum values for each pair consistently above 0.96.  
Moreover, their similarity margins are substantial, with the majority exceeding 0.6,  
indicating that the correct counterpart stands out clearly from all alternatives.

By contrast, new experts introduced in the addition and replacement scenarios do not exhibit a clear match.  
Their similarity distributions are comparatively uniform across victim experts,  
with very small margins, for example, 0.070 and 0.103 for the added \texttt{GTSRB} and \texttt{MNIST} experts in Fig.~\ref{fig:effectiveness-matrices}(e),  
and only 0.058 for the substituted \texttt{GTSRB} in (f).  
These weak margins indicate the absence of strong task-specific correspondence,  
ensuring that such new experts are not misattributed as belonging to the victim model.

\subsection{Robustness to Parametric Tampering}\label{sec:parametric}

Parametric tampering alters the experts' internal parameters  without changing the overall MoE architecture.  
We evaluate three representative operations: pruning, fine-tuning, and parameter permutation.  Table~\ref{tab:parametric_tampering} reports the final results.

\subsubsection{Pruning: \textit{\name} withstands aggressive sparsity}

Two pruning methods are tested and applied to all experts:  
1) magnitude-based pruning with 30\% sparsity ratio, and  
2) WANDA~\cite{wanda}, with sparsity ratios of 20\% to 50\%.

Table~\ref{tab:parametric_tampering} shows that while both \textit{\name} and \textit{REEF} remain stable under pruning, 
\textit{\name} consistently achieves higher similarity, staying above 0.92 and outperforming \textit{REEF} by about 19\%.  
In contrast, PCS and ICS baselines at both model and expert levels collapse to near zero.

\begin{figure}[!htb]
    \centering
    \includegraphics[width=0.9\linewidth]{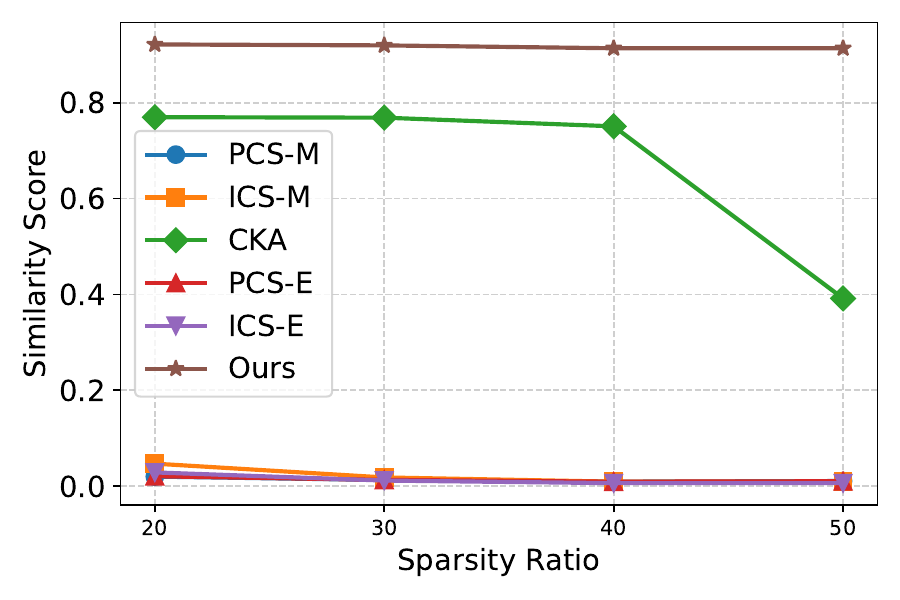}
    \caption{Fingerprint similarity after weights pruning using WANDA~\cite{wanda} with different sparsity ratio.}
    \label{fig:wanda}
\end{figure}

As illustrated in Fig.~\ref{fig:wanda}, \textit{REEF} begins to decline sharply once sparsity exceeds 40\%, 
dropping from about 0.8 to 0.4, 
whereas \textit{\name} remains above 0.9 even at 50\%.  
These results highlight the resilience of routing-based fingerprints under severe parameter pruning.

\begin{figure*}[!htb]
    \centering
    \subfigure[Expert Replacement]{\includegraphics[width=0.32\textwidth]{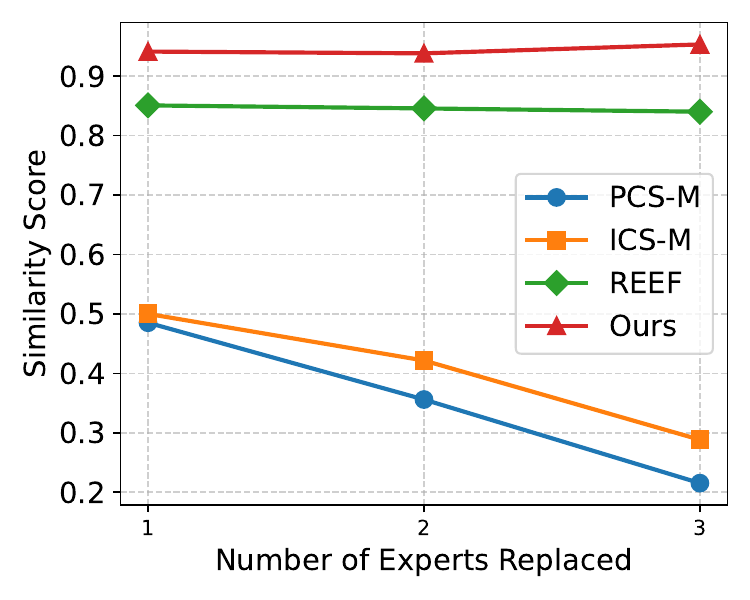}
    }
    \subfigure[Expert Deletion]{\includegraphics[width=0.32\textwidth]{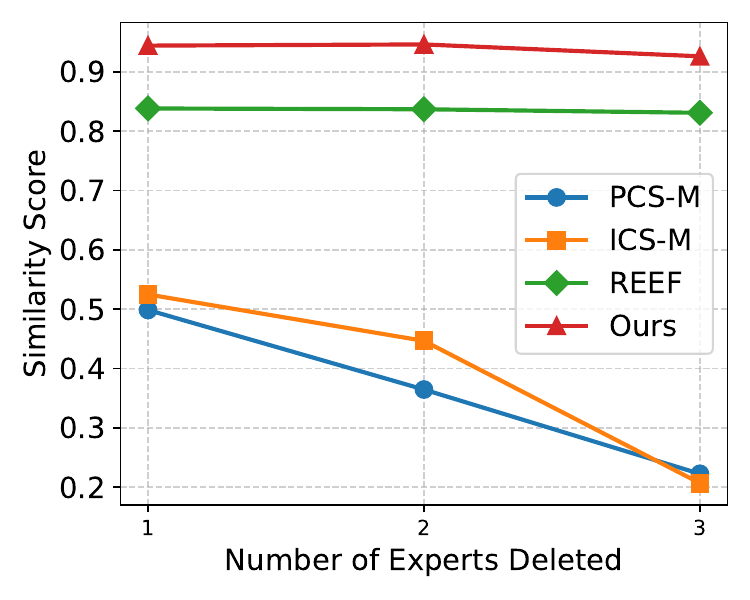}
    }
    \subfigure[Expert Addition]{\includegraphics[width=0.32\textwidth]{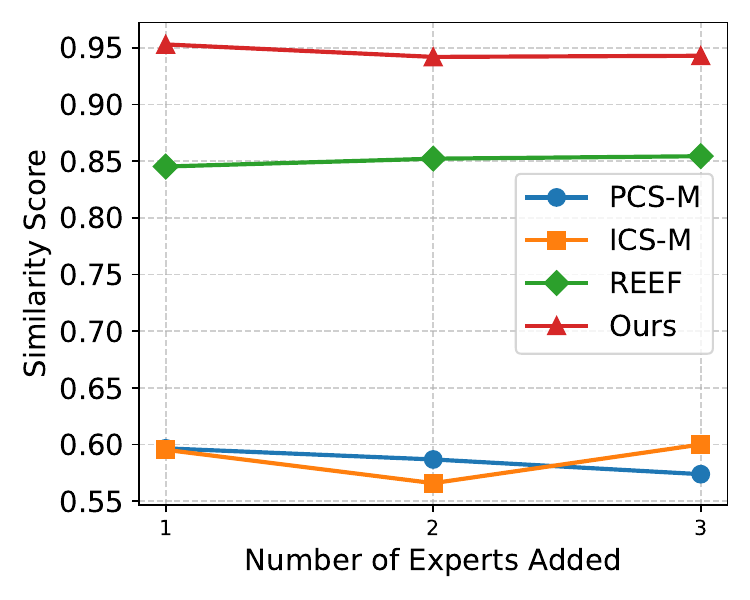}
    }
    \caption{Similarity scores under structural tampering. Each subfigure shows results for 1–3 experts being replaced, deleted, or added in the victim MoE. Our \textbf{\name} maintain consistently  stable and higher similarity for reused experts.}
    \label{fig:structural_tampering}
\end{figure*}

\subsubsection{Fine-tuning: \textit{\name} remains stable under extended training}

We fine-tune the victim model for 4,000 steps (\texttt{FT-4k}) and 8,000 steps (\texttt{FT-8k}), 
modeling post-merging adaptation where expert parameters continue to evolve.  
We use \texttt{FT-4k} and \texttt{FT-8k} to denote these two settings throughout the analysis.

As shown in Table~\ref{tab:parametric_tampering}, both model- and expert-level PCS and ICS baselines degrade noticeably as fine-tuning progresses, 
with \textit{PCS-M} dropping from 0.8396 at \texttt{FT-4k} to 0.7152 at \texttt{FT-8k}, and \textit{ICS-M} from 0.7138 to 0.6058.  
By contrast, \textit{REEF} and our \textit{\name} remain stable, 
yet \textit{\name} consistently achieves higher scores, reaching 0.958 and 0.956 compared to 0.851–0.843 for \textit{REEF}.  

\subsubsection{Permutation: \textit{\name} is invariant to parameter reordering}


In this setting, we permute all parameters of the merged MoE while preserving functional equivalence, 
so the model’s behavior remains unchanged even though weight alignment is disrupted.  
As a result, routing logits and activations are unaffected, allowing both \textit{\name} and \textit{REEF} to achieve perfect attribution with a similarity of 1.000, as shown in Table~\ref{tab:parametric_tampering}.  
In contrast, weight-based methods are severely disrupted: \textit{PCS-M} and \textit{PCS-E} collapse to near zero, 
while \textit{ICS-E}, which depends solely on low-rank and sparse task deltas, is highly sensitive to dimension reordering and thus suffers a sharp drop in similarity.  
Only \textit{ICS-M} retains a moderate value (0.574), because it aggregates parameters across all experts along with attention layers, 
providing partial robustness to permutation.

\subsection{Robustness to Structural Tampering}
\label{sec:structure_robustness}

\subsubsection{Settings}

Our structural tampering evaluation covers three operations: expert addition, deletion, and replacement.  
The corresponding suspects are built on the fine-tuned setting introduced in \S\ref{sec:parametric}.  
Because \textit{PCS-E} and \textit{ICS-E} measure similarities from expert-specific task deltas, 
their results here would merely replicate the fine-tuning analysis and are therefore omitted. 
We instead compare \textit{\name} with model-level baselines (\textit{PCS-M}, \textit{ICS-M}, \textit{REEF}), 
considering cases where the victim model of five experts is modified by replacing, adding, or deleting 1–3 experts trained on new tasks such as \texttt{GTSRB}, \texttt{MNIST}, and \texttt{DTD}.

\subsubsection{\textit{\name} is robust to the type and scale of structural tampering}

Across expert replacement, addition, and deletion, \textit{\name} consistently maintains high and stable similarity, remaining at or close to 0.9 across different tampering scales.
By contrast, \textit{PCS-M} and \textit{ICS-M} show clear declines as tampering intensifies across all three scenarios.  
The degradation is especially pronounced in replacement and deletion, where their similarity scores drop below 0.3 when three experts are affected.   This difference arises because replacement and deletion disrupt the victim model’s expert set to a much greater extent, 
whereas addition preserves all original experts and only augments the set with new ones, 
resulting in a relatively smaller degree of change.

\subsubsection{Routing is a stronger attribution signal than weight- or activation-based baselines.}

Although \textit{REEF} preserves relative robustness across structural tampering, 
its scores consistently lag behind those of \textit{\name}.  
Across scenarios and tampering scales, \textit{\name} exceeds \textit{REEF} by nearly 10\%.  
For example, in Fig.~\ref{fig:structural_tampering}(c), with three new experts added, 
\textit{\name} reaches about 0.95 while \textit{REEF} remains around 0.85.  
Comparable margins under replacement and deletion confirm that routing-based fingerprints capture richer and more reliable task-specific signals, 
making \textit{\name} a consistently stronger attribution method.

\subsection{Robustness to Varying Sample Sizes}

\begin{figure}[!htb]
    \centering
    \includegraphics[width=\linewidth]{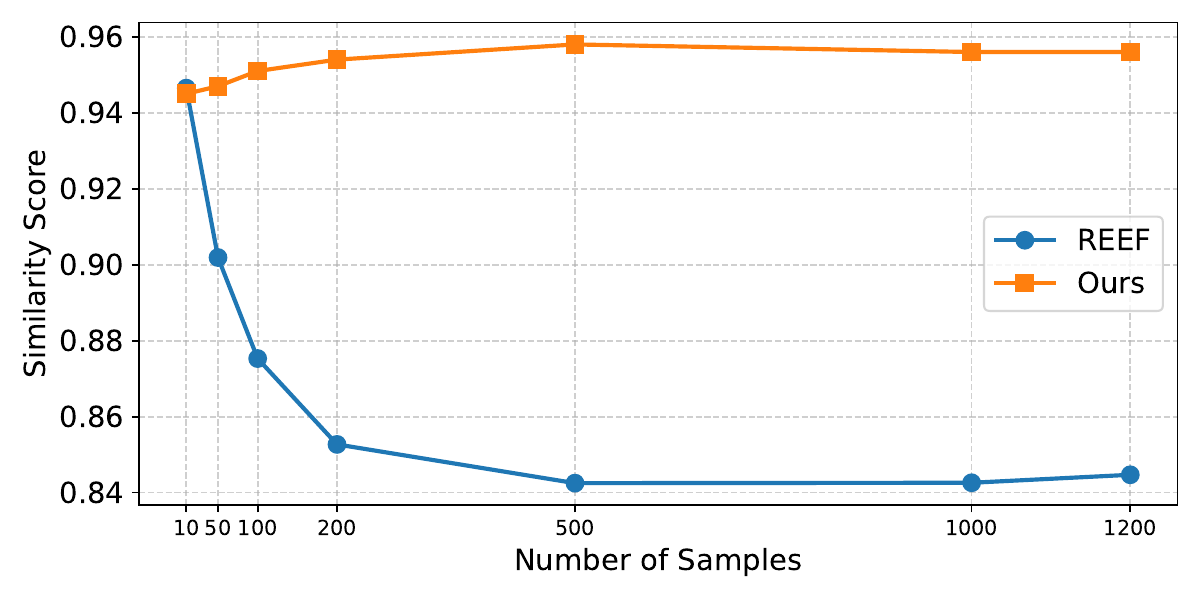}
    \caption{Fingerprint similarity under varying sample sizes. }
    \label{fig:sample_size}
\end{figure}

We further investigate how the number of probing samples influences the stability of fingerprint similarity. As shown in Fig.~\ref{fig:sample_size}, our routing-based method \textit{\name} consistently maintains high similarity scores above 0.94 across all sample sizes, and exhibits slight improvements as more samples are used. In contrast, the \textit{REEF} baseline demonstrates a sharp decline in similarity when the sample size increases, eventually stabilizing at a much lower level. We attribute this phenomenon to the growing complexity of intermediate activations in MoE models: as the number of samples increases, representations are influenced by a larger variety of experts, introducing noise that degrades the activation-based method’s ability to capture consistent expert behavior. These results highlight the sample efficiency and stability of our routing-based fingerprints, which remain reliable even with limited probing data.

\section{Conclusion}
In this work, we presented \textit{\name}, a routing-based fingerprinting framework for IP attribution in MoE-based model merging. 
To the best of our knowledge, this is the first systematic study addressing IP protection in dynamic, input-dependent merging scenarios. 
We proposed two complementary expert-level descriptors derived from routing behaviors: the \textit{RSF}, capturing activation intensity, and the \textit{RPF}, characterizing task-level specialization. 
Extensive experiments demonstrate that \textit{\name} consistently outperforms weight- and activation-based baselines, while remaining robust against both parametric and structural tampering operations. 
These findings establish routing-based fingerprints as a practical and effective solution for safeguarding IP in MoE-based model merging, 
and point toward future extensions to larger, more heterogeneous expert pools and multimodal MoE merging architectures.

\bibliography{aaai2026}

\end{document}